\documentclass[article]{jss}
\usepackage{bm}
\usepackage{amsmath,color}
\usepackage{amssymb,enumerate}
\usepackage{epsfig,graphicx}
\usepackage{amsthm}
\usepackage{booktabs}
\usepackage{multirow}
\usepackage{url}
\makeatletter
\g@addto@macro{\UrlBreaks}{\UrlOrds}
\makeatother
\graphicspath{{./Figures/}}

\def \calA{\mathcal{A}}
\def \calI{\mathcal{I}}
\def \RRR {\mathbb{R}} 

\def \veve{{\boldsymbol \varepsilon}}
\def \bbeta{\boldsymbol{\beta}}
\def \ggamma{\boldsymbol{\gamma}}

\def\bfm#1{\mathbf{#1}}
    
 \def\g{\bfm{g}} \def\h{\bfm{h}} \def\i2{\bfm{i}}

    \def\x{\bfm{x}} \def\y{\bfm{y}}

   \def\X{\bfm{X}} 
\def\Z{\bfm{Z}}

\def \one {\mathbf{1}}
\def \zero {\mathbf{0}}

\DeclareMathOperator*{\argmin}{arg\,min}

\newtheorem{remark}{Remark}

\author{Canhong Wen\\University of Science and \\ Technology of China\\ Sun Yat-sen University \And
	Aijun Zhang\\ The University of \\ Hong Kong \And
	Shijie Quan\\Sun Yat-sen University \And
        Xueqin Wang\\Sun Yat-sen University}
\title{\pkg{BeSS}: An \proglang{R} Package for Best Subset Selection in Linear, Logistic and CoxPH Models} 

\Plainauthor{Canhong Wen, Aijun Zhang, Shijie Quan, Xueqin Wang} 
\Plaintitle{\pkg{BeSS}: An R Package for Best Subset Selection in GLM and CoxPH Model} 
\Shorttitle{\pkg{BeSS}: Best Subset Selection} 

\Abstract{
We introduce a new \proglang{R} package, \pkg{BeSS},  for solving the best subset selection problem in linear, logistic and Cox's proportional hazard (CoxPH) models.
It utilizes a highly efficient active set algorithm based on primal and dual variables, and supports sequential and golden search strategies for best subset selection.
We provide a \proglang{C++}  implementation of the algorithm using \pkg{Rcpp} interface.
We demonstrate through numerical experiments based on enormous  simulation and real datasets that the new \pkg{BeSS} package has competitive performance compared to other \proglang{R} packages for best subset selection  purposes.
}
\Plainkeywords{best subset selection, primal dual active set, variable selection, R, C++, Rcpp} 

\Address{
  Canhong Wen\\
  Department of Statistics and Finance, School of Management\\
  University of Science and Technology of China\\
  230026  Hefei, AH, China\\
   E-mail: \email{wench@ustc.edu.cn}\\
  {\it and}  \\
  Department of Statistics, School of Mathematics\\
    Southern China Center for Statistical Science\\
  Sun Yat-sen University\\
  510275 Guangzhou,  GD, China\\

  Aijun Zhang\\
  Department of Statistics and Actuarial Science\\
  The University of Hong Kong\\
  Hong Kong, China\\
  E-mail: \email{ajzhang@hku.hk}\\

  Shijie Quan\\
  Department of Statistics, School of Mathematics\\
   Southern China Center for Statistical Science\\
  Sun Yat-sen University\\
  510275 Guangzhou,  GD, China\\

  Xueqin Wang\\
  Department of Statistics, School of Mathematics\\
  Southern China Center for Statistical Science\\
  Sun Yat-sen University\\
  510275 Guangzhou,  GD, China\\
  E-mail: \email{wangxq88@mail.sysu.edu.cn}\\
}

\begin{document}


\section{Introduction}

One of the main tasks of statistical modeling is to exploit the association between a response variable and multiple predictors. Linear model (LM), as a simple parametric regression model, is often used to capture linear dependence between response and predictors. The other two common models: generalized linear model (GLM) and Cox's proportional hazards (CoxPH) model, can be considered as the extensions of linear model, depending on the types of responses. Parameter estimation in these models can be computationally intensive when the number of predictors is large.  Meanwhile, Occam's razor is widely accepted as a heuristic rule for statistical modeling, which balances goodness of fit and model complexity. This rule leads to a relative small subset of important predictors.

The canonical approach to subset selection problem is  to choose $k$ out of $p$ predictors for each $k \in \{0,1,2,\dots,p\}$. This involves exhaustive search over all possible $2^p$ subsets of predictors, which is an NP-hard combinatorial optimization problem. 
To speed up, \cite{furnival1974regressions}  introduced a well-known branch-and-bound algorithm
with an efficient updating strategy for LMs, which was later implemented by \proglang{R} packages such as the \pkg{leaps} \citep{lumley2017leaps} and the \pkg{bestglm} \citep{mcleod2010bestglm}. Yet for GLMs, a simple exhaustive screen is undertaken in \pkg{bestglm}. When the exhaustive screening is not feasible for GLMs, fast approximating approaches have been proposed based on a genetic algorithm. For instance, \pkg{kofnGA}\citep{wolters2015a} implemented a genetic algorithm to search for a best subset of a pre-specified model size $k$, while \pkg{glmuti} \citep{calcagno2010glmulti} implemented a genetic algorithm to automatically select the best model for GLMs with no more than 32 covariates.
These packages can only deal with dozens of predictors but not high-dimensional data arising in modern statistics. Recently, \cite{bertsimas2016best} proposed a mixed integer optimization approach to find feasible best subset solutions for LMs with relatively larger $p$, which relies on certain third-party integer optimization solvers. Alternatively, regularization strategy is widely used to transform the subset selection problem into computational feasible problem. For example, \pkg{glmnet} \citep{friedman2010regularization, simon2011regularization} implemented a coordinate descent algorithm to solve the LASSO problem, which is a convex relaxation by replacing the cardinality constraint in best subset selection problem by the $L_1$ norm.

In this paper, we consider a primal-dual active set (PDAS) approach to solve the best subset selection problem for LM, GLM and CoxPH models. The PDAS algorithm for linear least squares problems was first introduced by \cite{ito2013variational} and later discussed by \cite{jiao2015primal},  \cite{Huang-et-al-2017} and \cite{ghilli2017monotone}. It utilizes an active set updating strategy  and fits the sub-models through use of complementary primal and dual variables.  We generalize the PDAS algorithm for general convex loss functions with the best subset constraint, and further extend it to support both sequential and golden section search strategies for optimal $k$ determination.  We develop a new package \pkg{BeSS} (BEst Subset Selection, \cite{wen2017bess}) in the \proglang{R} programming system \citep{R} with \proglang{C++} implementation of PDAS algorithms and memory optimized for sparse matrix output. This package is publicly available from the Comprehensive \proglang{R} Archive Network (CRAN) at \url{https://cran.r-project.org/package=BeSS}.
We demonstrate through enormous datasets that \pkg{BeSS} is efficient and stable for high dimensional data, and may solve best subset problems with $n$ in 1000s and $p$ in 10000s in just seconds on a single personal computer.

The article is organized as follows. In Section 2, we provide a general primal-dual formulation for the best subset problem that includes linear, logistic and CoxPH models as special cases. Section 3 presents the PDAS algorithms and related technical details. Numerical experiments with enormous simulations and real datasets are conducted in Section 4. We conclude with a short discussion in Section 5.

\section{Primal-dual formulation}\label{sec:2}
The best subset selection problem with the subset size $k$ is given by the following optimization problem:
\begin{equation}\label{eqn:best}
	\min_{\beta \in \mathbb{R}^p} \ l(\bbeta)  \quad \text{ s.t. } \quad \|\bbeta\|_0 = k,
\end{equation}
where $l(\bbeta)$ is a convex loss function of the model parameters $\bbeta \in \mathbb{R}^p$
and $k$ is an unknown positive integer.  The $L_0$ norm $\|\bbeta\|_0 =\sum_{j=1}^p |\beta_j|_0 = \sum_{j=1}^p 1_{\beta_j\neq0}$ counts the number of nonzeros in $\bbeta$. 


 It is known that the problem (\ref{eqn:best}) admits non-unique local optimal solutions, among which the coordinate-wise minimizers possess   promising properties. For a coordinate-wise minimizer $\bbeta^\diamond$, denote the vectors of gradient and Hessian diagonal  by
\begin{equation}\label{gh}
\g^\diamond = \nabla l (\bbeta^\diamond), \quad
\h^\diamond = \mbox{\rm diag}(\nabla^2 l (\bbeta^\diamond)),
\end{equation}
respectively.  For each coordinate $j=1,\ldots,p$, write $l_j(t) = l(\beta_1^\diamond, \dots, \beta_{j-1}^\diamond, t, \beta_{j+1}^\diamond, \dots, \beta_p^\diamond)$
while fixing the other coordinates.
Then the local quadratic approximation of  $l_j(t)$ around $\beta_j^\diamond$ is given by
\begin{equation}\label{eqn:quadapprox}
\arraycolsep=1.4pt\def\arraystretch{2.0}
  \begin{array}{l l}
l_j^Q(t) & = l_j(\beta_j^\diamond) +  g_j^\diamond (t-\beta_j^\diamond) + \frac{1}{2}  h_j^\diamond (t - \beta_j^\diamond)^2\\
&=  \dfrac{1}{2}  h_j^\diamond \left(t - \beta_j^\diamond +  g_j^\diamond/h_j^\diamond  \right)^2 + l_j(\beta_j^\diamond) -  \dfrac{1}{2}[g_j^\diamond] ^2/h_j^\diamond \\
&= \dfrac{1}{2}  h_j^\diamond \left(t - (\beta_j^\diamond + \gamma_j^\diamond) \right)^2 + l_j(\beta_j^\diamond) -  \dfrac{1}{2}[g_j^\diamond] ^2/h_j^\diamond,
  \end{array}
\end{equation}
which gives rise of an important quantity $\gamma_j^\diamond$  of the following scaled gradient form
\begin{equation}\label{gamma}
\gamma_j^\diamond = - g_j^\diamond/h_j^\diamond.
\end{equation}
Minimizing the objective function $l_j^Q(t)$ yields $t_j^*= \beta_j^\diamond +\gamma_j^\diamond$ for $j=1,\ldots,p$.

The constraint in  (\ref{eqn:best}) says that there are $p-k$ components of $\{t_j^*,  j=1,\ldots,p\}$ that would be enforced to be zero. To determine  which $p-k$ components, we consider the sacrifices  of $l_j^Q(t)$ when switching each $t_j^*$ from $ \beta_j^\diamond +\gamma_j^\diamond$ to $0$, which are given by
\begin{equation}\label{eqn:Delta}
\Delta_j^\diamond =\frac{1}{2} h_j^\diamond (\beta_j^\diamond +\gamma_j^\diamond)^2, \quad j=1,\ldots,p.
\end{equation}
Among all the candidates, we may enforce those $t_j^*$'s to zero 
if they contribute the {\em least total sacrifice} to the overall loss.
To realize this, let $\Delta_{[1]}^\diamond\geq \cdots \geq \Delta_{[p]}^\diamond$ denote the decreasing rearrangement of $\Delta_j^\diamond$ for $j=1,\ldots,p$, then truncate the ordered sacrifice vector at position $k$.
Combining the analytical result by (\ref{eqn:quadapprox}), we obtain that
\begin{equation}\label{eqn:hard}
 \beta_j^\diamond  = \left\{
  \begin{array}{l l}
    \beta_j^\diamond + \gamma_j^\diamond , & \text{ if }  \Delta_j^\diamond\geq \Delta_{[k]}^\diamond  \\
    0, & \text{ otherwise}.
  \end{array} \right. 
\end{equation}

In (\ref{eqn:hard}), we treat $\bbeta^\diamond = (\beta_1^\diamond, \ldots, \beta_p^\diamond)$ as primal variables, $\ggamma^\diamond = (\gamma_1^\diamond, \ldots, \gamma_p^\diamond)$ as dual variables, and $\bm\Delta^\diamond = (\Delta_1^\diamond, \ldots, \Delta_p^\diamond)$ as reference sacrifices.
Next we provide their explicit expressions for three important statistical models. 

\bigskip
\noindent \textbf{Case 1: Linear regression}.
Consider the LM $\y=\X\bbeta + \veve$ with design matrix $\X\in\RRR^{n\times p}$ and i.i.d. errors.  
Here $\X$ and $\y$ are standardized such that the intercept term is removed from the model and each column of $\X$ has $\sqrt{n}$ norm.

Take the loss function $l(\bbeta)=\frac{1}{2n}\|\y-\X\bbeta\|^2$. For $j=1,\ldots,p$, it is easy to obtain
\begin{equation}\label{eqn:gh:lm}
g_j^\diamond = \frac{1}{n} \X_{(j)}^T(\X\bbeta-\y), \quad h_j^\diamond= 1,
\end{equation}
where 
$\X_{(j)}$ denotes the $j$th column of $\X$, so
\begin{equation}\label{eqn:delta:lm}
\gamma_j^\diamond = \frac{1}{n} \X_{(j)}^\top(\y - \X\bbeta), \quad \Delta_j^\diamond =\frac{1}{2}(\beta_j^\diamond + \gamma_j^\diamond)^2.
\end{equation}

\medskip
\noindent \textbf{Case 2: Logistic regression}.
Consider the GLM
$$
\log(p(\x)/(1-p(\x))) = \beta_0 + \x^\top\bbeta,  \quad \x\in\RRR^p
$$
with $p(\x) = \mbox{\rm Prob}(Y=1|\x)$. Given the data $\big\{(\x_i, y_i)\big\}_{i=1}^n$ with binary responses $y_i\in\{0,1\}$,  the negative log-likelihood function is given by
\begin{equation}\label{eqn:loglike:glm}
l(\beta_0, \bbeta) = - \sum_{i=1}^n \Big\{ y_i(\beta_0 + \x_i^\top\bbeta) -  \log(1+\exp(\beta_0 + \x_i^\top\bbeta)) \Big\}.
\end{equation}
We give only the primal-dual quantities for $\bbeta\in\RRR^p$ according to the $L_0$ constraint in  (\ref{eqn:best}), while leaving $\beta_0$ to be estimated by unconstrained maximum likelihood method.  For $j=1,\ldots,p$,
\begin{equation}\label{eqn:gh:glm}
g_j^\diamond = -\sum_{i=1}^nx_{ij}(y_i - p_i^\diamond), \quad  h_j^\diamond = \sum_{i=1}^n x_{ij}^2p_i^\diamond (1-p_i^\diamond)
\end{equation}
where $p_i^\diamond ={\exp(\beta_0 +  \x_i^\top\bbeta^\diamond)}/{(1 + \exp(\beta_0  + \x_i^\top\bbeta^\diamond))}$  denotes the $i$-th predicted probability.  Then,
\begin{equation}\label{eqn:delta:glm}
\gamma_j^\diamond = \dfrac{\sum_{i=1}^n x_{ij} (y_i - p_i^\diamond) }{ \sum_{i=1}^n x_{ij}^2 p_i^\diamond(1-p_i^\diamond)}, \quad\Delta_j^\diamond =  \frac{1}{2}{\sum_{i=1}^n x_{ij}^2 p_i^\diamond(1-p_i^\diamond)} (\beta_j^\diamond+\gamma_j^\diamond)^2.
\end{equation}

%
%
%


\medskip
\noindent \textbf{Case 3:  CoxPH regression.} Consider the CoxPH model
$$
\lambda(t|\x) =  \lambda_0(t) \exp(\x^\top\bbeta), \quad \x\in\RRR^p
$$
with an unspecified baseline hazard $\lambda_0(t)$. Given the data $\{(T_i,\delta_i, \x_i): i=1,\dots,n\}$ with observations of survival time $T_i$ and censoring indicator $\delta_i$,
by the method of partial likelihood  \citep{cox1972regression},  the parameter $\bbeta$ can be obtained by minimizing the following convex loss
\begin{equation}\label{eqn:cox:like}
l(\bbeta) = - \sum_{i: \delta_i=1}\bigg(\x_i^\top\bbeta - \log\bigg(\sum_{i': T_{i'}\geq T_i} \exp(\x_{i'}^\top\beta) \bigg) \bigg).
\end{equation}
By writing $\omega_{i,i'}^\diamond = \exp(\x_{i'}^\top\bbeta^\diamond)/\sum_{i': T_{i'}\geq T_i} \exp(\x_{i'}^\top\bbeta^\diamond)$, it can be verified that
\begin{align}\label{eqn:gh:cox}
g_j^\diamond & =  - \sum_{i: \delta_i=1}\bigg(x_{ij} -  \sum_{i': T_{i'}\geq T_i} \omega_{i,i'}^\diamond x_{i'j}\bigg)\\
h_j^\diamond &  = \sum_{i: \delta_i=1} \sum_{i': T_{i'}\geq T_i} \omega_{i,i'}^\diamond\bigg(x_{i'j} -  \sum_{i': T_{i'}\geq T_i}
\omega_{i,i'}^\diamond x_{i'j}\bigg)^2
\end{align}
so that $\gamma_j^\diamond = -g_j^\diamond/h_j^\diamond$ and $\Delta_j^\diamond = \frac{1}{2}h_j^\diamond(\beta_j^\diamond + \gamma_j^\diamond)^2$ for $j=1,\ldots,p$.

\section{Active set algorithm}
For the best subset problem  (\ref{eqn:best}), define the active set $\calA = \{j: \beta_j  \neq 0\}$ with cardinality $k$ and the inactive set $\calI  = \{j: \beta_j = 0\}$ with cardinality $p-k$.
For  the coordinate-wise minimizer $\bbeta^\diamond$ satisfying (\ref{eqn:hard}), we have that
\begin{enumerate}[(C1)]
\item \label{itm:C1} $\beta_j^\diamond = 0$ when $j\in \calI$;
\item $\gamma_j^\diamond= 0$ when $j\in \calA $;\label{con:C2}
\item $\Delta_j^\diamond \geq \Delta_{j'}^\diamond$ whenever $j\in\calA$ and $j' \in\calI$.\label{con:C3}
\end{enumerate}
By (C\ref{itm:C1}) and (C2), the primal variables $\beta_j^\diamond$'s and the dual variables $\gamma_j^\diamond$'s have complementary supports. (C3) can be viewed as a local stationary condition.
These three conditions lay the foundation for the primal-dual active set algorithm we develop in this section.

Let $\calA$ be a candidate active set.  By (C1), we may estimate the $k$-nonzero primal variables by standard convex optimization:
\begin{equation}\label{eqn:exactmin}
\bm{\hat\beta}  = 
\argmin_{\bbeta_{\calI}= \zero} l(\bbeta), \quad\mbox{where } \calI = \calA^c.
\end{equation}
Given $\bm{\hat\beta}$,  the $\g$ and $\h$ vectors  (\ref{gh}) can be computed, with their explicit expressions derived for linear, logistic and CoxPH models in the previous section. The $\ggamma$ and $\bm\Delta$ vectors are readily obtainable by (\ref{gamma}), (\ref{eqn:Delta}) and (C2). Then we may check if (C3) is satisfied; otherwise, update the active and inactive sets by
\begin{equation}\label{eqn:act}
\calA\leftarrow\left\{j: \Delta_j \geq \Delta_{[k]}\right\} , \qquad \calI\leftarrow \left\{j: \Delta_j  < \Delta_{[k]} \right\}.
\end{equation}
This leads to the following iterative algorithm.

\noindent\rule{\textwidth}{1.5pt}
  \textbf{Algorithm 1} Primal-dual active set (PDAS) algorithm\\
\noindent\rule{\textwidth}{0.8pt}

\vspace{-0.3cm}
\begin{enumerate}
\item Specify the cardinality $k$ of the active set and the maximum number of iterations $m_{\max}$.
Initialize $\calA$ to be a random $k$-subset of $\{1,\ldots,p\}$ and  $\calI = \calA^c$.
\item For $m = 1,2,\dots, m_{\max}$, do
\begin{enumerate}[(2.a)]
\item Estimate $\bm{\hat\beta}$ by (\ref{eqn:exactmin});
\item Compute $\g, \h, \ggamma, \bm\Delta$;
\item Update $\calA, \calI$ by (\ref{eqn:act});
\item If $A$ is invariant, stop. 
\end{enumerate}
\item Output $\{\calA, \bm{\hat\beta}, \bm\Delta\}$.
\end{enumerate}
\vspace{-0.5cm}
\noindent\rule{\textwidth}{1.5pt}

\begin{remark}
The proposed PDAS algorithm is close to the primal-dual active set strategy first developed by  \cite{ito2013variational}, but different from their original  algorithm in two main aspects. First, our PDAS algorithm is derived from the quadratic argument  (\ref{eqn:quadapprox}) and it involves the second-order partial derivatives (i.e. Hessian diagonal $\h$). Second, our algorithm extends the original linear model setting to the general setting with convex loss functions.
\end{remark}

\subsection{Determination of optimal $k$}\label{sec:tuning}

The subset size $k$ is usually unknown in practice, thus one has to determine it in a data-driven manner. A heuristic way is using the cross-validation technique to achieve the best prediction performance. Yet it is time consuming to conduct the cross-validation method especially for high-dimensional data. An alternative way is to run the PDAS algorithm from small to large $k$ values, then identify an optimal choice according to some criteria, e.g., Akaike information criterion (\cite{akaike1974new}, AIC) and Bayesian information criterion (\cite{schwarz1978estimating}, BIC) and  extended BIC (\cite{chen2008extended, chen2012extended}, EBIC) for small-$n$-large-$p$ scenarios.
This leads to the sequential PDAS algorithm.

\noindent\rule{\textwidth}{1.5pt}
  \textbf{Algorithm 2} Sequential primal-dual active set (SPDAS) algorithm\\
  \vspace{-0.5cm}
\noindent\rule{\textwidth}{0.8pt}
\vspace{-0.3cm}
\begin{enumerate}
\item Specify the maximum size $k_{\max}$ of the active set, and initialize $\calA^0=\emptyset$. 

\item For $k = 1,2,\dots, k_{\max}$,  do

\hspace{0.5cm}    \parbox{0.9\textwidth}{%
Run \textbf{PDAS} with initial $\calA^{k-1}\cup \{j \in\calI^{k-1}: j \in \mbox{arg}\max \Delta_j^{k-1}\} $. Denote the output by $\{\calA^{k}, \bbeta^k, \bm\Delta^k\}$.
    }%

\item Output the optimal choice $\{\calA^*, \bbeta^*, \bm\Delta^*\}$ that attains the minimum AIC, BIC or EBIC.
\end{enumerate}
\noindent\rule{\textwidth}{1.5pt}

\begin{figure}[!h]
\centering
\includegraphics[width=5in]{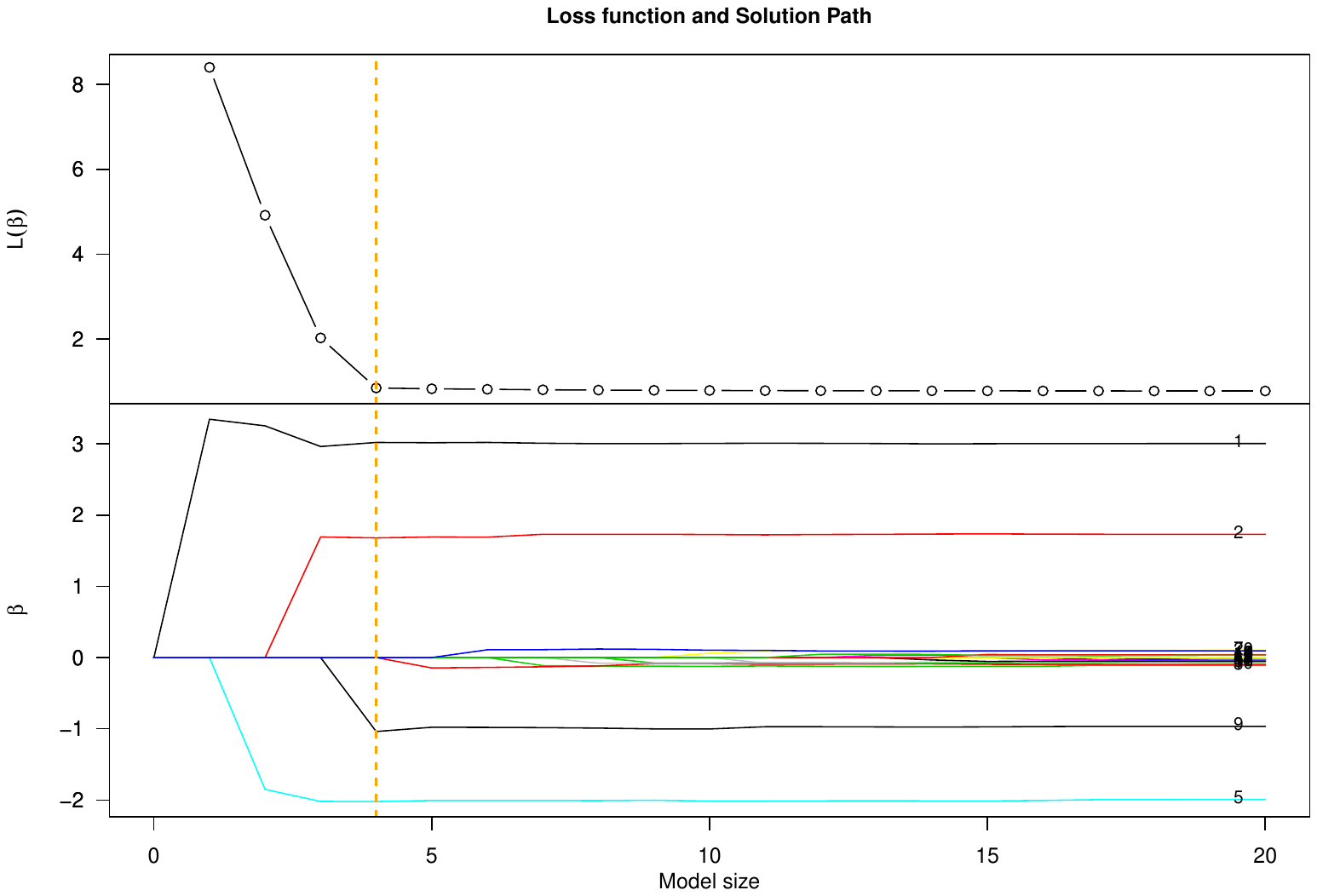}
\caption{Plot of the loss function  against the model complexity $k$ and solution path for each coefficients. The orange vertical dash line indicates number of true nonzero coefficients. }
\label{fig:loss}
\end{figure}

To alleviate the computational burden of determining $k$ as in SPDAS, here we provide an alternative method: the golden section search algorithm. We begin by plotting the loss function $l(\bbeta)$ as a function of $k$ for a simulated data from linear model with standard Gaussian error. The true coefficient $\bbeta=(3, 1.5, 0,0,-2,0,0,0,-1,0,\dots,0)$ and the design matrix $\X$ is generated as in Section \ref{sec:simulate} with $\rho=0.2$. From Figure~\ref{fig:loss}, it can be seen that the slope of the loss plot goes from steep to flat and there is an `elbow' exists near the true number of active set, i.e., $k=4$.
The solution path for the same data is presented at the bottom of Figure~\ref{fig:loss} for a better visualization on the relationship between loss function and coefficient estimation. When a true active predictor is included in the model, the loss function drops dramatically and the predictors already in the model adjust their estimates to be close to the true values. When all the active predictors are included in the model, their estimates would not change much as $k$ becomes larger.

Motivated by this interesting phenomenon, we develop a search algorithm based on the golden section method to determine the location of such an {\em elbow} in the loss function.  In this way, we can avoid to run the PDAS algorithm extensively for a whole sequential list. The golden section primal-dual active set (GPDAS) algorithm is summarized as follows.

\noindent\rule{\textwidth}{1.5pt}
  \textbf{Algorithm 3} Golden section primal-dual active set (GPDAS) algorithm \\
  \vspace{-0.5cm}
\noindent\rule{\textwidth}{0.8pt}
\vspace{-0.3cm}
\begin{enumerate}
\item Specify the number of maximum iterations $m_{\max}$, the maximum size $k_{\max}$ of the active set  and the tolerance $\eta \in (0,1)$. Initialize $k_L=1$, and $k_R=k_{\max}$.
\item For $m = 1,2,\dots, m_{\max}$,  do
\begin{enumerate}[(2.a)]
\item Run {PDAS} with $k = k_L$ and initial   $\calA^{m-1}_L \cup  \{j \in\calI^{m-1}_L: j \in \mbox{arg}\max (\Delta_L^{m-1})_{j}\} $. Output $\{\calA^{m}_L, \bbeta^m_L, \bm\Delta^m_L\}$.

\item Run {PDAS} with $k = k_R$ and initial $\calA^{m-1}_R \cup  \{j \in\calI^{m-1}_R: j \in \mbox{arg}\max (\Delta_R^{m-1})_{j}\} $. Output $\{\calA^{m}_R, \bbeta^m_R, \bm\Delta^m_R\}$.

\item Calculate $k_M = k_L + 0.618\times (k_R -k_L)$. Run {PDAS} with $k = k_M$ and initial  $\calA^{m-1}_M \cup  \{j \in\calI^{m-1}_M: j \in \mbox{arg}\max (\Delta_M^{m-1})_{j}\} $. Output $\{\calA^{m}_M, \bbeta^m_M, \bm\Delta^m_M\}$.
\item Determine whether $k_M$ is an  elbow  point:
\begin{itemize}
\item Run {PDAS} with $k = k_M-1$ and initial  $\calA^{m}_M $. Output $\{\calA^{m}_{M-}, \bbeta^m_{M-}, \bm\Delta^m_{M-}\}$.
\item Run {PDAS} with $k = k_M+1$ and initial  $\calA^{m}_M $. Output $\{\calA^{m}_{M+}, \bbeta^m_{M+}, \bm\Delta^m_{M+}\}$.
\item
If ${|l(\bbeta^{m}_M) - l(\bbeta^{m}_{M_-})|}>\eta{  | l(\bbeta^{m}_M)|}$ and ${|l(\bbeta^{m}_M) - l(\bbeta^{m}_{M_+})|}< \eta {  | l(\bbeta^{m}_M)|}/2$, then stop and denote $k_M$ as an elbow point, otherwise go ahead.
\end{itemize}
\item Update $k_L, k_R$  and $\calA ^{m}_L, \calA ^{m}_R$:
\begin{itemize}
\item If ${|l(\bbeta^{m}_M) - l(\bbeta^{m}_{L})|}>\eta{  | l(\bbeta^{m}_M)|}>{|l(\bbeta^{m}_R) - l(\bbeta^{m}_{L})|}$, then $k_R = k_M$, $\calA ^{m}_R = \calA ^{m}_M$;
\item If $\min \left\{{|l(\bbeta^{m}_M) - l(\bbeta^{m}_{L})|},{|l(\bbeta^{m}_R) - l(\bbeta^{m}_{L})|}\right\} > \eta{  | l(\bbeta^{m}_M)|}$, then $k_L = k_M$, $\calA ^{m}_L = \calA ^{m}_M$;
\item Otherwise, $k_R = k_M, \calA ^{m}_R = \calA ^{m}_M$ and $k_L = 1, \calA ^{m}_L = \emptyset $.

\end{itemize}
\item If $k_L= k_R - 1$, then  stop, otherwise $m=m+1$.
\end{enumerate}
\item Output $\{\calA^m_M, \bbeta^{m}_M, \bm\Delta^m_M\}$.
\end{enumerate}
\noindent\rule{\textwidth}{1.5pt}


\subsection{Computational details}
The proposed PDAS, SPDAS and SPDAS algorithms are much faster than existing methods reviewed in Section 1.  For the exhaustive methods like \pkg{leaps} and \pkg{bestglm}, they essentially deal with $\sum_{k=1}^{k_{\max}} C(p,k)$ sub-models in order to search for the best subset with size no more than $k_{\max}$.
It is infeasible even when $k_{\max}$ is moderate. That is why the greedy methods (e.g., \pkg{glmuti}) and the relaxed methods (e.g., \pkg{glmnet}) become popular.
Our proposed algorithms belong to the greedy methods and their computational complexity is discussed below.

In general, consider one iteration in step (2) of the PDAS algorithm with  a pre-specified $k$.
Denote by $N_{l}$ the computational complexity for solving $\bbeta$  on the active set; and
denote by $N_g$ and $N_h$ the computational complexity for calculating $\g$ and $\h$, respectively.  The calculation of $\ggamma$ in steps (2.b)-(2.c) costs $O((p-k)\max(N_h, N_g) )$, and the calculation of $\bm\Delta$ in steps (2.b)-(2.c) costs $O(pN_h)$. Then the overall cost of one iteration is $O(\max(N_{l}, pN_h, (p-k)N_g))$.

The total number of iterations of the PDAS algorithm could depend on the signal-to-noise ratio, the dimensionality $p$, and the sparsity level $k$. The algorithm may usually converge in finite steps (otherwise capped by $m_{\max}$).  Denote by $N_{\text{P}}$ the complexity for each run of the PDAS algorithm, then the total complexity of the SPDAS and GPDAS algorithms are $O(k_{\max} \times N_{\text{P}})$ and $O(\log(k_{\max}) \times N_{\text{P}})$, respectively.

\noindent \textbf{Case 1: Linear regression.}  Since $\h=\one$, $N_h = 0$.  The matrix vector product in the computation of $\h$ takes $O(n)$ flops.  For the least squares problem on the active set, we use Cholesky factorization to obtain the estimate, which leads to $N_{l}= O(\max(nk^2, k^3))$. Thus the total cost of one iteration in step (2) is ${O}(\max(n k^2, k^3, n(p-k)))$, and the overall cost of the PDAS algorithm is the same since the number of iterations is often finite.

In particular, if the true coefficient vector is sparse with $ k \ll p$ and $n =O(\log(p))$, the cost of the PDAS algorithm is $O(np)$, a linear time with respective to the size $p$.
With an unknown $k$, we can choose an appropriate $k_{\max}$ value, e.g., $k_{\max}=n/\log(n)$, to speed up the SPDAS and GPDAS algorithms. Their costs become $O(n^2p/\log(n))$ and  $O(np\log(n/\log(n)))$, respectively. These rates are comparable with the sure independence screening procedure \citep{fan2008sure} in handling ultrahigh-dimensional data. In fact, even if the true coefficient vector is not sparse, we could use a conjugate gradient \citep{golub2012matrix} algorithm with a preconditioning matrix to achieve a similar computational rate.


\noindent \textbf{Case 2: Logistic regression.} It costs $O(p)$ flop to compute the predicted probabilities $p_i$'s. Thus $N_g=O(np)$ and $N_h=O(np)$. We use the iteratively reweighted least squares (IRLS) 
for parameter estimation on the active set. The complexity of each IRLS step is the same as that of the least squares, so $N_{l}= O(N_I \max(nk^2, k^3))$ with $N_I$ denoting the finite number of IRLS iterations. The total cost of one iteration in step (2) is ${O}(\max(np^2, nk^2N_I, k^3N_I))$.


\noindent \textbf{Case 3: CoxPH regression.} It costs $O(np)$ flops to compute $\omega_{i,i'}$'s. Assume the censoring rate is $c$, then $N_g=O(n^3p(1-c))$ and $N_h=O(n^3p(1-c))$.
Like the \code{coxph} command from the \pkg{survival} package, we adopt the standard Newton-Raphson algorithm for the maximum partial likelihood estimation on the active set. Its difficulty arises in the computation of the inverse of the Hessian matrix, which is full and dense.  The Hessian matrix has $k^2$ entries and it requires $O(n^3 k(1-c))$ flops for the computation of each entry. The matrix inversion costs $O(k^3)$ via Gauss-Jordan elimination or Cholesky decomposition.  Hence, for each Newton-Raphson iteration, the updating equation requires $O(\max(n^3 k^3(1-c), k^3))$ flops. We may speed up the algorithm by replacing the Hessian matrix with its diagonal, which reduces the computational complexity per updating to $O(\max(n^3 k^2(1-c), k^3))$. Denote by $N_{nr}$ the number of Newton-Raphson iterations, then $N_{l}= O(N_{nr} \max(n^3 k^2(1-c), k^3))$ and the total cost of one iteration in step (2) is ${O}(\max(n^3p^2(1-c), n^3 k^2(1-c)N_{nr}, k^3N_{nr})). $ 

\subsection{R package}
We have implemented the active set algorithms described above into an \proglang{R} package called \pkg{BeSS} (BEst Subset Selection), which is publicly available from the CRAN at \url{https://cran.r-project.org/package=BeSS}. The package is implemented in \proglang{C++} with memory optimized using sparse matrix output
and it can be called from \proglang{R} by a user-friendly interface.

The package contains two main functions, i.e., \code{bess.one} and \code{bess},  for solving the best subset selection problem with or without specification of $k$. In \code{bess}, two options are provided to determine the optimal $k$: one is based on the SPDAS algorithm with criteria including AIC, BIC and EBIC; the other is based on the GPDAS algorithm. The function \code{plot.bess} generates plots of loss functions for the best sub-models for each candidate $k$, together with solution paths for each predictor. We also include functions \code{predict.bess} and \code{predict.bess.one} to make prediction on the new data.

\section{Numerical examples}
In this section we compare the performance of our new \pkg{BeSS} package to other well-known packages for best subset selection: \pkg{leaps}, \pkg{bestglm} and \pkg{glmulti}. We also include \pkg{glmnet} as an approximate subset selection method and use the default cross-validation method to determine an optimal tuning parameter. All parameters use the default values of the corresponding main functions in those packages unless otherwise stated. In presenting the results of \pkg{BeSS}, \code{bess.seq} represents \code{bess} with argument \code{method = "sequential"} and \code{bess.gs} represents \code{bess} with argument \code{method = "gsection"}, two different ways to determine the optimal parameter $k$. In \code{bess.seq}, we use AIC for examples with $n\geq p$ and EBIC
for examples with $n<p$. 
We choose $k_{\max} = \min(n/2, p)$ for linear models and $k_{\max} = \min(n/\log(n), p)$ for logistic and CoxPH models.

 All the \proglang{R} codes are demonstrated in Section 4.3.  All computations are carried out on a 64-bit Intel machine with a single 3.30 GHz CPU and 4 GB of RAM.

\subsection{Simulation data}\label{sec:simulate}
We demonstrate the practical application of our new \pkg{BeSS} package on synthetical data under both low and high dimensional settings. For the low-dimensional data, \pkg{BeSS} has comparable performance with other state-of-the-art methods. For the high-dimensional data, while most state-of-the-art methods become incapable to deal with them, \pkg{BeSS} still performs fairly well.  For an instance, \pkg{BeSS} is scalable enough to identify the best sub-model over all candidates efficiently in seconds or a few minutes when the dimension $p = 10000$.

We compare the performances of different methods in three aspects. The first aspect is the run time in seconds (Time). The second aspect is the selection performance in terms of true positive (TP) and false positive (FP) numbers, which are defined by the numbers of true relevant and true irrelevant variables among the selective predictors. The third aspect is the predictive performance on a held out test data of size 1000. For linear regression, we use the relative mean squares error (MSE) as defined by $\|\X\hat\bbeta - \X\bbeta^*\|_2/\|\X\bbeta^*\|_2$. For logistic regression, we calculate the classification accuracy by the average number of observations being correctly classified.  For CoxPH regression, we compute the median time on the test data, then derive the area under the receiver operator characteristic curve (i.e., AUC) using nearest neighbor estimation method as in \cite{heagerty2000time}.

We generate the design matrix $\X$ and the underlying coefficients $\bbeta$ as follows. The design matrix $\X$ is generated with $\X_{(j)} = \Z_j + 0.5\times(\Z_{j-1} + \Z_{j+1}), ~j=1,\dots, p$, where $\Z_0=\zero, \Z_{p+1}=\zero$ and $\{\Z_j, j=1,\dots,p\}$ were  i.i.d. random samples drawn from standard Gaussian distribution and subsequently normalized to have $\sqrt{n}$ norm. The true coefficient $\bbeta^*$ is a vector with $q$ nonzero entries uniformly distributed in $[b, B]$, with  $b$ and $B$ to be specified.  In the simulation study, the sample size is fixed to be $n=1000$. For each scenario, 100 replications are conducted .

\textbf{Case 1: Linear regression.} For each $\X$ and $\bbeta^*$, we generate the response vector $\y = \X\bbeta^* + \sigma \epsilon$, with  $\epsilon \sim \mathcal{N}(0,1)$. We set $b=5\sigma \sqrt{2\log(p)/n}, ~B = 100b$ and $\sigma=3$.
Different choices of $(p, q)$ are taken to cover both the low-dimensional cases $(p = 20, 30, \text{or }40, \ q=4)$ and the high-dimensional cases ($p = 100, 1000,  \text{or }10000,\ q= 40$). For \code{glmulti}, we only present the result for $p=20$ and $p=30$ since it can only deal with at most 32 predictors. Since \code{leaps} and \code{bestglm} cannot deal with high-dimensional case,  we only report the results of \code{glmnet}, \code{bess.seq} and \code{bess.gs}. The results are summarized in Table~\ref{tab:sim:lm}.

In the low-dimensional cases,  the performances of all best subset selection methods are comparable in terms of prediction accuracy and selection consistency. However, the regularization method \code{glmnet} has much higher MSE and lower FP, which suggests that LASSO incurs bias in the coefficient estimation. In terms of computational times, both \code{bess.seq} and \code{bess.gs} have comparable performance with \code{glmnet}, which cost much less run times than the state-of-the-art methods. Unlike \code{leaps}, \code{bestglm} and \code{glmulti}, the run times of \code{bess.seq} and \code{bess.gs} remain fairly stable across different dimensionality.

In the high-dimensional cases, both  \code{bess.seq} and \code{bess.gs} work quite well and they have similar performance in prediction and variable selection. Furthermore, their performances become better as $p$ and $q$ increase (from left to right in Table~\ref{tab:sim:lm}). On the other hand, \code{glmnet} has higher FP as $p$ increases. In particular, when $p=10000$ and only $40$ nonzero coefficients are involved, the average TP equals $40$  and the average FP is less than 3.06. In contrast,  the average FP of \code{glmnet} increases to 30.  As for the computational issues,  both  \code{bess.seq} and \code{bess.gs} seem to grow at a linear rate of $p$, but \code{bess.gs} offers speedups by factors of 2 up to 10 and more.

\begin{table}[!h]
\centering
\begin{tabular}{lccccc}
\toprule
  Low-dimensional & & Method & $p=20$ & $p=30$ & $p=40$ \\
\cmidrule(r){2-6}
& Time & \code{leaps} & 0.00(0.01) &  0.39(0.13) & 58.79(28.78) \\
 && \code{bestglm}& 0.02(0.01) &  0.51(0.15) & 69.39(32.27) \\
 && \code{glmulti}& 11.91(2.60) &  18.41(4.13) & --- \\
 && \code{glmnet} & 0.08(0.02) & 0.09(0.02) & 0.08(0.01) \\
 && \code{bess.seq}& 0.18(0.01) &  0.23(0.02) &  0.25(0.03) \\
 && \code{bess.gs} & 0.16(0.01) &  0.18(0.02) &  0.17(0.02) \\
     \cmidrule(r){2-6}
&MSE & \code{leaps} & 1.91(0.83) & 2.18(0.81) & 2.44(1.15) \\
   &($\times 10^{-2}$) & \code{bestglm} & 1.91(0.83) & 2.18(0.81) & 2.44(1.15) \\
   && \code{glmulti}& 1.87(0.72) & 2.16(0.79) & --- \\
   && \code{glmnet} & 3.90(1.30) & 3.51(1.23) & 3.51(1.37) \\
   && \code{bess.seq} & 1.93(0.82) & 2.12(0.76) & 2.43(1.21) \\
   && \code{bess.gs} & 2.14(2.45) & 2.06(1.78) & 2.80(3.37) \\
     \cmidrule(r){2-6}
&TP & \code{leaps}& 3.97(0.17) & 3.99(0.10) & 3.97(0.17) \\
   && \code{bestglm} & 3.97(0.17) & 3.99(0.10) & 3.97(0.17) \\
   && \code{glmulti}& 3.99(0.10) &  4.00(0.00) & --- \\
   && \code{glmnet} & 3.96(0.20) & 3.97(0.17) & 3.95(0.22) \\
   && \code{bess.seq} & 3.96(0.20) & 3.91(0.35) & 3.84(0.44) \\
   && \code{bess.gs} & 3.78(0.42) & 3.73(0.51) & 3.63(0.61) \\
     \cmidrule(r){2-6}
&FP & \code{leaps} & 2.37(1.83) & 3.92(2.39) & 5.53(2.66) \\
   && \code{bestglm} & 2.37(1.83) & 3.92(2.39) & 5.53(2.66) \\
   && \code{glmulti}& 2.29(1.63) &  4.15(2.29) & --- \\
   && \code{glmnet} & 0.73(0.80) & 0.82(0.83) & 0.78(1.10) \\
   && \code{bess.seq} & 3.75(4.25) & 4.98(5.80) & 7.59(8.64) \\
   && \code{bess.gs} & 1.35(2.94) & 4.31(6.93) & 5.42(8.74) \\
   \midrule
 High-dimensional & & Method & $p=100$ & $p=1000$ & $p=10000$ \\
\cmidrule(r){2-6}
 & Time & \code{glmnet} &  0.16(0.03) &  1.77(0.09) & 14.82(1.73) \\
& & \code{bess.seq} & 1.29(0.09) &  74.54(1.33) & 137.04(13.80) \\
& & \code{bess.gs} & 0.53(0.12) &   3.72(0.41) &  12.87(2.89) \\
          \cmidrule(r){2-6}
 & MSE &  \code{glmnet}& 1.42(0.18) &  2.51(0.28) &  2.47(0.22) \\
   & ($\times 10^{-2}$)   & \code{bess.seq} & 1.65(0.41) & 1.20(0.62) & 0.70(0.23) \\\
   && \code{bess.gs} & 1.33(0.29) & 0.98(0.37) & 1.00(0.35) \\
          \cmidrule(r){2-6}
&TP & \code{glmnet} & 39.74(0.54) & 39.80(0.45) & 39.75(0.46) \\
   && \code{bess.seq} & 35.30(2.17) & 38.72(1.29) & 39.53(0.70) \\
   && \code{bess.gs} & 35.78(2.12) & 39.43(0.88) & 39.58(0.71) \\
          \cmidrule(r){2-6}
&FP & \code{glmnet} & 15.45(3.65) & 12.73(5.50) & 29.82(11.91) \\
   && \code{bess.seq} & 27.15(10.66) &  4.92(6.99) &  0.32(1.92) \\
   && \code{bess.gs} & 28.86(8.90) &  1.51(2.53) &  3.06(3.84) \\
   \bottomrule
\end{tabular}\label{tab:sim:lm}
\caption{Simulation results for linear regression. Time stands for run time (CPU seconds), MSE stands for Mean Squared Error, TP stands for true positive number and FP stands for false positive number. The number of true nonzero coefficients is $q=4$ for low-dimensional cases and $q=40$ for high-dimensional cases. }
\end{table}

\textbf{Case 2: Logistic regression.} For each $\x$ and $\bbeta^*$,  the binary response is generated by $y = \text{Bernoulli}(\mbox{\rm Prob}(Y=1))$, where $\mbox{\rm Prob}(Y=1) = \exp(\x^\top\bbeta^*)/(1+\exp(\x^\top\bbeta^*))$. The range of nonzero coefficients are set as $b=10 \sqrt{2\log(p)/n},\ B = 5b$. Different choices of $p$ are taken to cover both the low-dimensional cases $(p = 8, 10, \text{ or }12)$ and the high-dimensional cases $(p = 100, 1000,  \text{ or } 10000)$. The number of true nonzero coefficients is chosen to be $q=4$ for low-dimensional cases and $q=20$ for high-dimensional cases. Since \code{bestglm} is based on complete enumeration, it may be used for low-dimensional cases yet it becomes computationally infeasible for high dimensional cases.

The simulation results are summarized in Table~\ref{tab:sim:logistic}.  When $p$ is small, both \code{bess.seq} and \code{bess.gs} have comparable performance with  \code{bestglm}, \code{glmulti} and \code{glmnet},  but have considerably faster speed in computation than \code{bestglm} and \code{glmulti}. In the high-dimensional cases, we see that all three methods perform very well in terms of accuracy and TP. Yet both \code{bess.seq} and \code{bess.gs} have much smaller FP than \code{glmnet}. Among them, the run time for \code{bess.gs} is around a quarter of that for \code{bess.seq} and is similar to that for \code{glmnet}.

\begin{table}[!h]
\centering
\begin{tabular}{lccccc}
\toprule
  Low-dimensional & & Method& $p=8$ & $p=10$ & $p=12$ \\
\cmidrule(r){2-6}
&Time  & \code{bestglm} &  1.83(0.15) &  7.55(0.26) & 28.35(1.93) \\
 && \code{glmulti} &  2.08(0.11)  &  13.91(2.43) &  21.61(4.54) \\
   && \code{glmnet} &  0.49(0.07)  &  0.56(0.09) &  0.63(0.17) \\
   && \code{bess.seq} &   0.70(0.33) &  0.79(0.35) &  0.78(0.52) \\
   && \code{bess.gs} &   0.52(0.20) &  0.78(1.14) &  0.65(0.23) \\
     \cmidrule(r){2-6}
   &Acc    & \code{bestglm} &  0.949(0.012) & 0.950(0.013) & 0.950(0.011) \\
    && \code{glmulti} &  0.949(0.012)  &  0.950(0.013) &  0.950(0.011) \\
   && \code{glmnet} &  0.949(0.013) & 0.951(0.013) & 0.950(0.011) \\
   && \code{bess.seq} & 0.949(0.012) & 0.950(0.013) & 0.950(0.011) \\
   && \code{bess.gs} & 0.948(0.013) & 0.951(0.012) & 0.949(0.013) \\
     \cmidrule(r){2-6}
&TP & \code{bestglm} & 3.99(0.10) & 4.00(0.00) & 3.99(0.10) \\
 && \code{glmulti} &  3.99(0.10)  &  4.00(0.00) &  4.00(0.00) \\
 &  & \code{glmnet}& 4.00(0.00) & 4.00(0.00) & 4.00(0.00) \\
  & & \code{bess.seq}& 3.96(0.20) & 3.95(0.30) & 3.91(0.32) \\
   & & \code{bess.gs}& 3.87(0.37) & 3.87(0.42) & 3.89(0.40) \\
     \cmidrule(r){2-6}
&FP & \code{bestglm} & 0.73(0.85) & 1.02(1.05) & 1.41(1.44) \\
 && \code{glmulti} &  0.73(0.85)  &  1.02(1.05) &  1.37(1.20) \\
   && \code{glmnet} & 1.62(0.96) & 2.07(1.16) & 2.83(1.44) \\
   && \code{bess.seq} & 1.77(1.59) & 2.19(2.20) & 2.39(2.40) \\
   && \code{bess.gs} & 0.15(0.41) & 0.31(0.93) & 0.64(1.57) \\
  [1ex]
   \midrule
    [0.1ex]
  High-dimensional & & Method& $p=100$ & $p=1000$ & $p=10000$ \\
\cmidrule(r){2-6}
&Time & \code{glmnet} &   4.75(0.89) &   4.38(0.49) &  17.01(0.24) \\
   && \code{bess.seq} &  43.99(7.42) &  54.85(4.46) & 108.66(2.47) \\
   && \code{bess.gs} &   7.34(2.10) &  11.46(1.81) &  22.43(2.16) \\
     \cmidrule(r){2-6}
&Acc & \code{glmnet} & 0.969(0.006) & 0.945(0.009) & 0.922(0.011) \\
   && \code{bess.seq} & 0.963(0.012) & 0.972(0.011) & 0.979(0.006) \\
   && \code{bess.gs} & 0.970(0.010) & 0.976(0.008) & 0.978(0.009) \\
      \cmidrule(r){2-6}
&TP & \code{glmnet} & 19.96(0.20) & 19.97(0.17) & 19.79(0.52) \\
   && \code{bess.seq} & 16.50(2.38) & 19.34(1.23) & 19.92(0.34) \\
   && \code{bess.gs} & 18.62(1.15) & 19.81(0.49) & 19.82(0.61) \\
     \cmidrule(r){2-6}
&FP & \code{glmnet} &  34.59(4.74) & 122.82(19.80) & 222.77(43.63) \\
   && \code{bess.seq} &   5.61(3.37) &   1.82(2.03) &   0.49(0.67) \\
   && \code{bess.gs} &   3.16(2.46) &   0.95(1.34) &   0.54(0.92) \\
   \bottomrule
\end{tabular}\label{tab:sim:logistic}
\caption{Simulation results for logistic regression.  Time stands for run time (CPU seconds), Acc stands for classification accuracy, TP stands for true positive number and FP stands for false positive number.  The number of true nonzero coefficients is $q=4$ for low-dimensional cases and $q=20$ for high-dimensional cases. }
\end{table}

\textbf{Case 3: CoxPH regression.} For each $\x$ and $\bbeta^*$, we generate data from the CoxPH model with hazard rate $\lambda(t|\x) = \exp(\x^\top\bbeta^*)$. The ranges of nonzero coefficients are set same as those in logistic regression, i.e., $b=10 \sqrt{2\log(p)/n}, ~B = 5b$. Different choices of $p$ are taken to cover both the low-dimensional cases $(p = 8, 10, \text{ or }12)$ and the high-dimensional cases $(p = 100, 1000,  \text{ or } 10000)$. The number of true nonzero coefficients is chosen to be $q=4$ for low-dimensional cases and $q=20$ for high-dimensional cases. Since \code{glmulti} cannot handle more than 32 predictors, we only report the low dimensional result for \code{glmulti}.

The simulation results  are summarized in Table~\ref{tab:sim:cox}. Our findings about \code{bess.seq} and \code{bess.gs} are similar to those for the logistic regression.

\begin{table}[!h]
\centering
\begin{tabular}{lccccc}
\toprule
  Low-dimensional & & Method& $p=8$ & $p=10$ & $p=12$ \\
\cmidrule(r){2-6}
&Time & \code{glmulti} &  1.53(0.06)& 10.11(1.75) &  15.20(2.86) \\
   && \code{glmnet} &  1.07(0.20)  &  1.09(0.20) &  1.16(0.23) \\
   && \code{bess.seq} &   0.42(0.20) &  0.49(0.23) &  0.52(0.22) \\
   && \code{bess.gs} &   0.35(0.15) &  0.46(0.19) &  0.51(0.18) \\
     \cmidrule(r){2-6}
   &AUC & \code{glmulti} &  0.973(0.012)  &  0.972(0.010) &  0.974(0.010) \\
   && \code{glmnet} &  0.973(0.012) & 0.972(0.010) & 0.974(0.010) \\
   && \code{bess.seq} & 0.973(0.012) & 0.972(0.010) & 0.974(0.010) \\
   && \code{bess.gs} & 0.972(0.012) & 0.972(0.010) & 0.974(0.011) \\
     \cmidrule(r){2-6}
&TP& \code{glmulti} &  4.00(0.00)  &  3.99(0.10) &  4.00(0.00) \\
 &  & \code{glmnet}& 4.00(0.00) & 4.00(0.00) & 4.00(0.00) \\
  & & \code{bess.seq}& 4.00(0.00) & 4.00(0.00) & 4.00(0.00) \\
   & & \code{bess.gs}& 3.89(0.35) & 3.96(0.20) & 3.99(0.10) \\
     \cmidrule(r){2-6}
&FP& \code{glmulti} &  0.60(0.77)  &  1.06(1.17) &  1.14(1.21) \\
   && \code{glmnet} & 1.17(1.01) & 1.56(1.04) & 1.82(1.14) \\
   && \code{bess.seq} & 1.62(1.69) & 1.98(2.25) & 2.38(2.69) \\
   && \code{bess.gs} & 0.11(0.35) & 0.04(0.20) & 0.06(0.37) \\
  [1ex]
   \midrule
    [0.1ex]
  High-dimensional & & Method& $p=100$ & $p=1000$ & $p=10000$ \\
\cmidrule(r){2-6}
&Time  & \code{glmnet} &  16.61(1.90) & 297.01(62.83) & 832.69(73.26) \\
  & & \code{bess.seq}&  20.57(1.77) &  72.53(2.58) & 233.53(11.94) \\
  & & \code{bess.gs}&   4.86(1.59) &  15.36(1.69) &  63.23(7.21) \\
     \cmidrule(r){2-6}
&AUC   & \code{glmnet} & 0.993(0.005) & 0.992(0.006) & 0.991(0.007) \\
  & & \code{bess.seq} & 0.993(0.005) & 0.992(0.006) & 0.991(0.007) \\
  & & \code{bess.gs} & 0.990(0.008) & 0.992(0.006) & 0.991(0.007) \\
     \cmidrule(r){2-6}
&TP  & \code{glmnet} & 20.00(0.00) & 20.00(0.00) & 20.00(0.00) \\
  & & \code{bess.seq} & 18.06(1.67) & 19.70(0.70) & 20.00(0.00) \\
  & & \code{bess.gs} & 17.09(2.03) & 19.93(0.33) & 19.99(0.10) \\
     \cmidrule(r){2-6}
&FP  & \code{glmnet} &  41.26(4.10) & 245.82(19.41) & 541.13(34.33) \\
  & & \code{bess.seq} &  11.80(9.25) &   1.64(3.78) &   0.02(0.14) \\
  & & \code{bess.gs} &  13.65(11.84) &   0.19(0.60) &   0.05(0.22) \\
   \bottomrule
\end{tabular}\label{tab:sim:cox}
\caption{Simulation results for CoxPH regression. Time stands for run time (CPU seconds),  AUC stands for the integrated time-dependent area under the curve, TP stands for true positive number and FP stands for false positive number. The number of true nonzero coefficients is $q=4$ for low-dimensional cases and $q=20$ for high-dimensional cases. }
\end{table}

\subsection{Real data}\label{sec:real}

We also evaluate the performance of the \pkg{BeSS} package in modeling several real data sets. Table~\ref{tab:data} lists these instances and their descriptions. All datasets
are saved as \proglang{R} data objects and available online with this publication.

\begin{table}[!h]
\centering
\renewcommand{\arraystretch}{1.2}
\begin{tabular}{lccll}
\toprule
Dataset  & $n$ & $p$ & Type    & Data source   \\
\midrule
\code{prostate} & 97   & 9       & Continuous & \proglang{R} package \pkg{ElemStatLearn}\\
\code{SAheart}  & 462  & 8       & Binary  & \proglang{R} package \pkg{ElemStatLearn}   \\
\code{trim32}   & 120  & 18975   & Continuous &\cite{scheetz2006regulation} \\
\code{leukemia} & 72   & 3571    & Binary    & \proglang{R} package \pkg{spikeslab} \\
{\code{gravier}}   & {168}  &{2905}   & {Binary}  & \href{https://github.com/ramhiser/datamicroarray/wiki/Gravier-(2010)}{https://github.com/ramhiser/}\\
\code{er0}      & 609  & 22285   & Survival   & \href{https://www.ncbi.nlm.nih.gov/geo/}{https://www.ncbi.nlm.nih.gov/geo/}\\
   \bottomrule
\end{tabular}\label{tab:data}
\caption{Description for the real data sets.  Here $n$ denotes the number of observations, $p$ denotes the number of predictors, and `Type' denotes the type of response.}
\end{table}

We randomly split the data into a training set with two-thirds observations and a test set with remaining observations. Different best subset selection methods are used to identify the best sub-model. For each method, the run time in seconds (Time) and the size of selected model (MS) are recorded. We also include measurements of the predictive performance on test data according to the metrics as in Section~\ref{sec:simulate}. For reliable evaluation, the aforementioned procedure is replicated for 100 times.

The modeling results are displayed in Table~\ref{tab:real}. Again in low-dimensional cases, \code{bess} has comparable performance with the state-of-art algorithms (branch-and-bound algorithm for linear models and complete enumeration algorithm and genetic algorithm for GLMs). Besides, \code{bess.gs} has comparable run time with \code{glmnet} and is considerably faster than \code{bess.seq} especially in high-dimensional cases.

\begin{table}[!h]
\centering
\renewcommand{\tabcolsep}{0.1cm}
\begin{tabular}{lccccccc}
  \toprule
  Data &Method & \code{leaps} & \code{bestglm}& \code{glmulti} & \code{glmnet} &\code{bess.seq} & \code{bess.gs}\\
  \midrule
\code{prostate} & Time & 0.00(0.01) & 0.01(0.01)&0.61(0.05) & 0.07(0.01) &  0.22(0.01) &  0.22(0.01)  \\
& PE & 0.61(0.14) & 0.61(0.14)& 0.61(0.14) & 0.65(0.19) & 0.60(0.13) & 0.60(0.14)  \\
& MS  & 4.27(1.11) & 4.25(1.12)& 4.25(1.12)& 3.58(0.87) & 4.29(1.17) & 6.11(0.87)   \\
\cmidrule(r){1-8}
\code{SAheart} &   Time & --- & 1.58(0.07)& 4.03(0.53)& 0.13(0.01) &  0.27(0.04) &  0.26(0.04)  \\
& Acc &  ---  & 0.72(0.03)&0.72(0.03) & 0.70(0.04)& 0.72(0.03) & 0.72(0.03)  \\
& MS    & --- & 5.68(0.98)& 5.68(0.98)& 4.61(0.84)& 5.68(0.99) & 6.29(1.09)  \\
\cmidrule(r){1-8}
\code{trim32} & Time &  --- &  --- &  ---  & 3.23(0.15)&  1.95(0.53) &  1.08(0.19)  \\
& PE &  ---  &  --- &  --- & 0.01(0.01)& 0.01(0.01) & 0.01(0.00)  \\
& MS    &  ---  &  --- &  ---  & 24.89(11.79)& 1.60(0.62) & 7.82(2.26)  \\
\cmidrule(r){1-8}
\code{leukemia} & Time &  ---  &  ---  &  --- & 0.38(0.01)&  1.74(0.77) &  1.14(0.53)  \\
& Acc &  ---  &  ---  &  ---  & 0.93(0.05)& 0.90(0.06) & 0.91(0.06)  \\
& MS    &  ---  &  ---  &  ---  & 11.76(4.40)& 1.54(0.77) & 2.00(0.00)  \\
\cmidrule(r){1-8}
\code{gravier} & Time &  ---  &  --- &  ---   & 0.68(0.03)&  6.64(4.09) &  2.93(2.50)  \\
& Acc &  ---  &  ---  &  ---  & 0.71(0.07)& 0.72(0.06) & 0.72(0.06)  \\
& MS   &  ---  &  --- &  ---  & 10.83(7.39)& 9.23(1.05) & 10.80(2.47)  \\
   \cmidrule(r){1-8}
\code{er0} & Time &  ---  &  ---  &  ---  & 154.97(15.75)&  184.51(86.15) &  55.20(22.07)  \\
& AUC &  ---  &  ---  &  ---  & 0.52(0.04)& 0.53(0.05) & 0.60(0.05)  \\
& MS    &  ---  &  --- &  ---  & 3.06(7.35)& 1.02(0.14) & 56.85(6.90)  \\
       \bottomrule
\end{tabular}\label{tab:real}
\caption{Results for the real data sets. Time stands for run time (CPU seconds),  MS stands for the size of selected model. PE stands for mean prediction error in linear model; Acc stands for classification accuracy in logistic regression model; AUC stands for the integrated time-dependent area under the curve in CoxPH regression model. }
\end{table}

\subsection{Code demonstration}
We demonstrate how to use the package \pkg{BeSS} on a synthesis data as discussed in Section~\ref{sec:tuning} and a real data in Section~\ref{sec:real}. Firstly, load \pkg{BeSS} and generate data with the \code{gen.data} function.

\begin{Sinput}
R> require("BeSS")
R> set.seed(123)
R> Tbeta <- rep(0, 20)
R> Tbeta[c(1, 2, 5, 9)] <- c(3, 1.5, -2, -1)
R> data <- gen.data(n = 200, p = 20, family = "gaussian", beta = Tbeta,
+    rho = 0.2, sigma = 1)				
\end{Sinput}

We may call the \code{bess.one} function to solve the best subset selection problem with a specified cardinality. Then we can \code{print} or \code{summary} the \code{bess.one} object. While the \code{print} method allows users to obtain a brief summary of the fitted model, the \code{summary} method presents a much more detailed description.

\begin{Sinput}
R> fit.one <- bess.one(data$x, data$y, s = 4, family = "gaussian")
R> print(fit.one)
\end{Sinput}
\begin{Soutput}
 	Df         MSE         AIC         BIC        EBIC
  4.0000000   0.8501053 -24.4790117 -11.2857422  12.6801159
\end{Soutput}
\begin{Sinput}
R> summary(fit.one)
\end{Sinput}
\begin{Soutput}
----------------------------------------------------------------------
    Primal-dual active algorithm with maximum iteration being 15

    Best model with k = 4 includes predictors:

       X1        X2        X5        X9
 3.019296  1.679419 -2.021521 -1.038276

    log-likelihood:    16.23951
    deviance:         -32.47901
    AIC:              -24.47901
    BIC:              -11.28574
    EBIC:              12.68012
----------------------------------------------------------------------
\end{Soutput}

The estimated coefficients of the fitted model can be extracted by using the \code{coef} function, which provides a sparse output with the control of argument \code{sparse = TRUE}. It is recommended to output a non-sparse vector when \code{bess.one} is used,  and to output a sparse matrix when \code{bess} is used.

\begin{Sinput}
R> coef(fit.one, sparse = FALSE)
\end{Sinput}
\begin{Soutput}
(intercept)          X1          X2          X3          X4          X5
-0.07506287  3.01929556  1.67941924  0.00000000  0.00000000 -2.02152109
         X6          X7          X8          X9         X10         X11
 0.00000000  0.00000000  0.00000000 -1.03827568  0.00000000  0.00000000
        X12         X13         X14         X15         X16         X17
 0.00000000  0.00000000  0.00000000  0.00000000  0.00000000  0.00000000
        X18         X19         X20
 0.00000000  0.00000000  0.00000000
\end{Soutput}

To make  prediction on new data, the \code{predict} function can be used as follows.

\begin{Sinput}
R> pred.one <- predict(fit.one, newdata = data$x)
\end{Sinput}

To extract the selected best model, we provide the \code{lm}, \code{glm}, or \code{coxph} type of object named
the \code{bestmodel} in the fitted \code{bess.one} object. 
Users could \code{print}, \code{summary} or \code{predict} this \code{bestmodel} object just like working with classical regression modeling. This would be helpful for statistical analysts who are familiar with \code{lm}, \code{glm}, or \code{coxph} functions.

\begin{Sinput}
R> bm.one <- fit.one$bestmodel
R> summary(bm.one)
\end{Sinput}
\begin{Soutput}
Call:
lm(formula = ys ~ xbest)

Residuals:
     Min       1Q   Median       3Q      Max
-2.54220 -0.63600 -0.04702  0.64100  3.11518

Coefficients:
            Estimate Std. Error t value Pr(>|t|)
(Intercept) -0.07506    0.06603  -1.137    0.257
xbestX1      3.01930    0.06715  44.962   <2e-16 ***
xbestX2      1.67942    0.06577  25.535   <2e-16 ***
xbestX5     -2.02152    0.06577 -30.735   <2e-16 ***
xbestX9     -1.03828    0.06313 -16.446   <2e-16 ***
---
Signif. codes:  0 '***' 0.001 '**' 0.01 '*' 0.05 '.' 0.1 ' ' 1

Residual standard error: 0.9338 on 195 degrees of freedom
Multiple R-squared:  0.9566,	Adjusted R-squared:  0.9557
F-statistic:  1075 on 4 and 195 DF,  p-value: < 2.2e-16
\end{Soutput}

In practice when the best  subset size is unknown, we have to determine the optimal choice of such sub-model size. The function \code{bess} provides two options: \code{method = "sequential"} corresponds to the SPDAS algorithm, and \code{method = "gsection"} corresponds to the GPDAS algorithm. Next we illustrate the usage of \code{bess} in the \code{trim32} data. We first load the data into the environment and show that it has 18975 variables, a much larger number compared with the sample size 120.

\begin{Sinput}
R> load("trim32.RData")	
R> dim(X)	
\end{Sinput}
\begin{Soutput}
[1]   120 18975
\end{Soutput}

Below is an example of running \code{bess} with argument \code{method = "sequential", epsilon = 0} and other argument being default values. We use the \code{summary} function to give a summary of the fitted \code{bess} object.

\begin{Sinput}
R> fit.seq <- bess(X, Y, method="sequential", epsilon = 0)		
R> summary(fit.seq)
\end{Sinput}
\begin{Soutput}
----------------------------------------------------------------------------
   Primal-dual active algorithm with tuning parameter determined by
   sequential method

   Best model determined by AIC includes 25 predictors with AIC = -890.9282

   Best model determined by BIC includes 25 predictors with BIC = -821.2409

   Best model determined by EBIC includes 2 predictors with EBIC = -561.2689
-----------------------------------------------------------------------------
\end{Soutput}

As in the \code{bess.one}, the \code{bess} function outputs an \code{lm} type of object \code{bestmodel} associated with the selected best model. Here the \code{bestmodel} component outputs the largest fitted model since we did not use any early stopping rule as shown in the argument \code{epsilon = 0}.

\begin{Sinput}
R> bm.seq <- fit.seq$bestmodel		
R> summary(bm.seq)
\end{Sinput}
\begin{Soutput}
Call:
lm(formula = ys ~ xbest)

Residuals:
      Min        1Q    Median        3Q       Max
-0.039952 -0.012366 -0.001078  0.011401  0.075677

Coefficients:
                   Estimate Std. Error t value Pr(>|t|)
(Intercept)        5.618703   0.407769  13.779  < 2e-16 ***
xbest1368348_at   -0.089394   0.014563  -6.139 1.97e-08 ***
xbest1370558_a_at -0.122228   0.010712 -11.410  < 2e-16 ***
xbest1372548_at   -0.179410   0.012085 -14.846  < 2e-16 ***
xbest1377032_at   -0.062936   0.016733  -3.761 0.000294 ***
xbest1382223_at    0.497858   0.023655  21.047  < 2e-16 ***
xbest1388491_at    0.266606   0.021538  12.378  < 2e-16 ***
xbest1388657_at   -0.085292   0.015030  -5.675 1.53e-07 ***
xbest1389122_at   -0.101926   0.015317  -6.655 1.88e-09 ***
xbest1390269_at    0.106434   0.012130   8.774 7.40e-14 ***
xbest1378024_at   -0.123666   0.017614  -7.021 3.40e-10 ***
xbest1378552_at   -0.049578   0.010397  -4.768 6.77e-06 ***
xbest1379586_at   -0.066086   0.013526  -4.886 4.22e-06 ***
xbest1379772_at   -0.096651   0.010166  -9.507 2.05e-15 ***
xbest1379933_at    0.186271   0.015806  11.785  < 2e-16 ***
xbest1380696_at    0.028347   0.006882   4.119 8.19e-05 ***
xbest1380977_at    0.104704   0.018148   5.769 1.01e-07 ***
xbest1382392_at   -0.033764   0.005830  -5.791 9.21e-08 ***
xbest1384690_at   -0.083789   0.013985  -5.991 3.80e-08 ***
xbest1385015_at    0.131036   0.011803  11.102  < 2e-16 ***
xbest1385032_at    0.100631   0.012171   8.268 8.73e-13 ***
xbest1385395_at   -0.139164   0.010919 -12.745  < 2e-16 ***
xbest1385673_at    0.071119   0.011828   6.013 3.46e-08 ***
xbest1392605_at   -0.051400   0.008229  -6.246 1.21e-08 ***
xbest1394502_at    0.020363   0.006134   3.320 0.001283 **
xbest1398128_at   -0.084070   0.012728  -6.605 2.36e-09 ***
---
Signif. codes:  0 '***' 0.001 '**' 0.01 '*' 0.05 '.' 0.1 ' ' 1

Residual standard error: 0.02241 on 94 degrees of freedom
Multiple R-squared:  0.981,	Adjusted R-squared:  0.976
F-statistic: 194.5 on 25 and 94 DF,  p-value: < 2.2e-16
\end{Soutput}

Alternatively, we might use criteria like AIC to select the best model among a sequential list of candidate models. As shown above, the output of the \code{bess} function includes the AIC, BIC and EBIC values for best subset selection. Since the \code{trim32} data is high dimensional, we opt to use the EBIC criterion to determine the optimal model size. Then we run the \code{coef} function to extract the coefficients in the \code{bess} object and output the nonzero coefficients of the selected model.

\begin{Sinput}
R> K.opt.ebic <- which.min(fit.seq$EBIC)
R> coef(fit.seq)[, K.opt.ebic][which(coef(fit.seq)[, K.opt.ebic]!=0)]
\end{Sinput}
\begin{Soutput}
 (intercept) 1382223_at 1388491_at
   0.8054785  0.5715478  0.3555834
\end{Soutput}

We can also run the \code{predict} function for a given \code{newdata}. The argument \code{type} specifies which criteria is used to select the best fitted model.

\begin{Sinput}
R> pred.seq <- predict(fit.seq, newdata = data$x, type="EBIC")
\end{Sinput}

\begin{figure}[!h]
\centering
\includegraphics[width=5in]{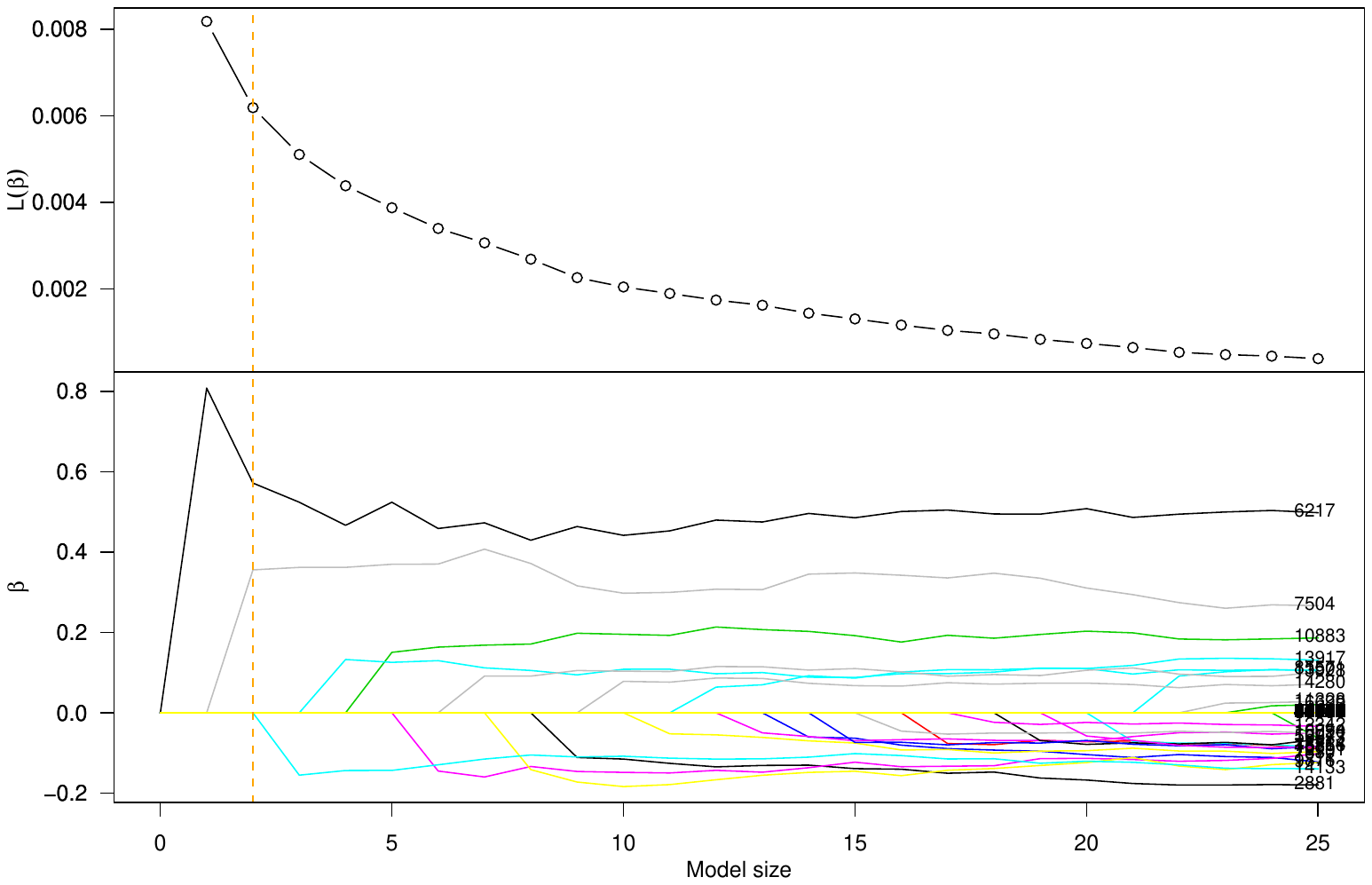}
\caption{Best subset selection results for the \code{trim32} data with \code{bess.seq}. The optimal $k$ value is determined by EBIC, which is indicated by a orange vertical dashed line.}
\label{fig:seq}
\end{figure}

The \code{plot} routine provides the loss function plot for the sub-models with different $k$ values, as well as solution paths for each predictor. It also adds a vertical dashed line to indicate the optimal $k$ value as determined by EBIC. Figure~\ref{fig:seq} shows the result from the following \proglang{R} code.

\begin{Sinput}
R> plot(fit.seq, type = "both", breaks = TRUE, K = K.opt.ebic)
\end{Sinput}

Next we call the function \code{bess} with argument \code{method = "gsection"} to perform the GPDAS algorithm. At each iteration, it outputs the split information. 

\begin{Sinput}
R> fit.gs <- bess(X, Y, family = "gaussian", method = "gsection",
R+		epsilon = 1e-2)
\end{Sinput}
\begin{Soutput}
1-th iteration s.left:1 s.split:16 s.right:25
2-th iteration s.left:1 s.split:10 s.right:16
3-th iteration s.left:1 s.split:7 s.right:10
4-th iteration s.left:1 s.split:5 s.right:7
5-th iteration s.left:5 s.split:6 s.right:7
\end{Soutput}

From the above code, we know that the best selected model has 6 predictors and the algorithm ends at the 5{\it th} iteration. To show more information about the best selected model, we may extract \code{fit.gs$bestmodel} and present its summary information via the S3 method \code{summary}.

\begin{Sinput}
R> bm.gs <- fit.gs$bestmodel
R> summary(bm.gs)
\end{Sinput}
\begin{Soutput}
Call:
lm(formula = ys ~ xbest)

Residuals:
      Min        1Q    Median        3Q       Max
-0.114598 -0.036829 -0.007365  0.041804  0.161688

Coefficients:
                  Estimate Std. Error t value Pr(>|t|)
(Intercept)        1.41219    0.50792   2.780 0.006363 **
xbest1368316_at   -0.16680    0.03479  -4.794 5.02e-06 ***
xbest1372248_at    0.22120    0.05579   3.965 0.000129 ***
xbest1373887_at    0.27947    0.05707   4.897 3.27e-06 ***
xbest1387160_at   -0.12456    0.02989  -4.168 6.05e-05 ***
xbest1389910_at    0.54459    0.07220   7.543 1.25e-11 ***
xbest1381978_a_at -0.16091    0.03454  -4.658 8.77e-06 ***
---
Signif. codes:  0 '***' 0.001 '**' 0.01 '*' 0.05 '.' 0.1 ' ' 1

Residual standard error: 0.05926 on 113 degrees of freedom
Multiple R-squared:  0.8405,	Adjusted R-squared:  0.8321
F-statistic: 99.28 on 6 and 113 DF,  p-value: < 2.2e-16
\end{Soutput}

%
%
%
%

Running the \code{coef} function directly on the \code{bess} object returns a sparse matrix as shown below. The last column corresponds to the best fitted coefficients.

\begin{Sinput}
R> beta <- coef(fit.gs, sparse = TRUE)
R> class(beta)
\end{Sinput}
\begin{Soutput}
[1] "dgCMatrix"
attr(,"package")
[1] "Matrix"
\end{Soutput}
\begin{Sinput}
R> beta[, ncol(beta)][which(beta[, ncol(beta)]!=0)]
\end{Sinput}
\begin{Soutput}
 (intercept)   1368316_at   1372248_at   1373887_at   1387160_at
   1.4121869   -0.1668030    0.2211982    0.2794672   -0.1245576
  1389910_at 1381978_a_at
   0.5445936   -0.1609133
\end{Soutput}

Based on the \code{fit.gs}, we can predict for the new data via the \code{predict} function as follows.

\begin{Sinput}
R> pred.gs <- predict(fit.gs, newdata = X)
\end{Sinput}

\section{Discussion}
In this paper, we introduce a primal dual active set (PDAS) algorithm for solving the best subset selection problem under the general convex loss setting. The PDAS algorithm identifies the best sub-model with a pre-specified model size via a primal-dual formulation on feasible solutions. To determine the best sub-model over different model sizes, both sequential search and golden section search are proposed, i.e., SPDAS and GPDAS algorithms. We find that the GPDAS algorithm is especially efficient and accurate in selecting variables for high-dimensional and sparse data.

The proposed algorithms are implemented with \proglang{C++} through the new  \pkg{BeSS} package in the \proglang{R} statistical environment. Package \pkg{BeSS} provides \proglang{R} users with a new and flexible way to carry out best subset selection for LM, GLM and CoxPH models. It allows us to identify the best sub model efficiently (usually in seconds or a few minutes) even when the number of predictors is extremely large, say $p \approx 10000$, based on a standard personal computer. In both simulation and real data examples, it was shown that the \pkg{BeSS} package is highly efficient compared to other state-of-the-art methods.


\section*{Acknowledgments}
We are grateful to the anonymous referees for valuable comments that lead to the improvement of the current paper. Wen's research is partially supported by NSFC(11801540), the Natural Science Foundation of Guangdong (2017A030310572), the Fundamental Research Funds for the Central Universities (WK2040170015, WK2040000016). Zhang's research is partially supported by Basic Research Seed Fund (201611159250) and Big Data Project Fund of The University of Hong Kong. Wang's research is partially supported by  NSFC(11771462), The National Key Research and Development Program of China(2018YFC1315400), and The Key Research and Development Program of Guangdong, China(2019B020228001).


\bibliographystyle{jss}
\bibliography{jss3186}

\end{document}